\documentclass[pra,amsmath,amssymb,twocolumn]{revtex4-1}
\usepackage{graphicx}
\usepackage{subfigure}
\usepackage{adjustbox}
\usepackage{bm}
\usepackage{color}
\usepackage{braket}
\usepackage{standalone}
\usepackage{multirow}
\usepackage{tikz}
\usepackage{mathrsfs}
\usepackage{dsfont}
\usepackage{times}

\usepackage[colorlinks,bookmarks=true,citecolor=blue,linkcolor=blue,urlcolor=blue]{hyperref}

\begin{document}

\title{Disorder-driven phase transitions in bosonic fractional quantum Hall liquids}
\author{Chao Han and Zhao Liu}
\affiliation{Zhejiang Institute of Modern Physics, Zhejiang University, Hangzhou 310027, China}
\date{\today}

\begin{abstract}
We investigate the disorder-driven phase transitions in bosonic fractional quantum Hall liquids at filling factors $f=1/2$ and $f=1$ in the lowest Landau level. We use the evolution of ground-state entanglement entropy, fidelity susceptibility, and Hall conductance with increasing disorder strength to identify the underlying phase transitions. The critical disorder strengths obtained from these different quantities are consistent with each other, validating the reliability of our numerical calculations based on exact diagonalization. At $f=1/2$, we observe a clear transition from the bosonic Laughlin state to a trivial insulating phase. At $f=1$, we identify a direct phase transition from the non-Abelian bosonic Moore-Read state to a trivial insulating phase, although some signs of a disorder-induced intermediate fractional quantum Hall phase were recently reported for the $f=5/2$ fermionic cousin. 
\end{abstract}
\maketitle

\section{Introduction}\label{introduction}
The phase transition is one of the most classical and long-standing problem in condensed matter physics. While Landau's symmetry-breaking theory gains great success in describing phase transitions, the discovery of integer~\cite{Klitzing80} and the more complex, interaction-induced fractional quantum Hall (FQH) effects~\cite{TSG82} in two-dimensional electron gases (2DEGs) penetrated by strong magnetic fields clearly indicates the existence of novel topological phases~\cite{Wen90} beyond the symmetry-breaking paradigm. Since the 1980s, various FQH states, their competing phases, and the transitions between them have been extensively studied~\cite{Laughlin83,Haldane83,Jain89,MooreRead91,Morf98,Stormer99,McDonald96,Ardonne99,Pan03,Gervais04,Jolicoeur07,Papic10,Archer13,Geraedts15,Zhao15,Peterson15,Pakrouski15,Thiebaut15,Gils09,Samkharadze16}.

Disorder is a ubiquitous ingredient that may drive phase transitions in FQH systems. Although weak disorder is responsible for the characteristic plateaus of Hall conductance when the FQH effect occurs, sufficiently strong disorder can destroy FQH phases by closing the spectral and mobility gaps. Unfortunately, due to the exponentially large Hilbert space of the underlying many-body system as well as the breaking of spatial symmetry, it is challenging to study in the microscopic level how disorder affects FQH phases. In the past decades, only a few numerical studies tackle this problem for fermionic FQH systems~\cite{Sheng03,Xin05,Friedman2011,Zhao16,Zhao17,Wei19}. By tracking the evolution of the ground-state energy gap, Hall conductance, and entanglement entropy as a function of disorder strength, disorder-driven transitions from Abelian and non-Abelian FQH phases to trivial phases were identified in microscopic models. In particular, for electrons at filling $f=5/2$ where numerical simulations support either the non-Abelian fermionic Moore-Read (MR) state or its particle-hole (PH) conjugate anti-MR state as the ground state in the zero-disorder limit~\cite{MooreRead91,Morf98,Rezayi00,Levin07,Lee07,Peterson08,Pakrouski15,Rezayi17}, it was found that enhancing disorder with a finite correlation length might first drive the system into an intermediate FQH phase before completely ruining the topological order~\cite{Wei19}. This observation may provide deep insight into the nature of the mysterious $f=5/2$ FQH effect which is still under debate~\cite{Son15,Xin16,Zucker16,Banerjee18,Mishmash18,Mross18,Chong18}.  

Most studies of FQH physics focus on electrons because of the direct relevance to 2DEG experiments. However, it has been proposed that bosons in rapidly rotating atomic gases or optical lattices can also form FQH states, where an effective magnetic field is created by rotation or laser beams~\cite{Cooper01,Nicolas03,Rezayi05,Sorensen05,Nicolas07,Cooper08,Fetter09,Gunnar09,Cooper13}. From the viewpoint of the wave function, there is a correspondence between the fermionic and bosonic Read-Rezayi series~\cite{Read99,Cooper01}, including the Laughlin~\cite{Laughlin83} and MR states, in which 
one can obtain a bosonic state from a fermionic one by simply removing a proper Jastrow factor to ensure the correct statistics. This fact immediately raises several interesting questions. How does disorder affect bosonic FQH states? Is the disorder-driven physics for bosonic FQH states similar to the corresponding fermionic cases? In particular, is there a disorder-induced intermediate FQH phase between the bosonic MR state and the trivial strong-disorder limit? 

To investigate these problems, in this paper we study the effect of disorder on bosonic FQH liquids at filling fractions $f=1/2$ and $f=1$ in the lowest Landau level (LLL). For contact-interacting bosons at $f=1/2$, we observe a direct phase transition from the bosonic Laughlin state to a trivial insulator, which is identified by the special behavior of the ground-state entanglement entropy, fidelity susceptibility, and Hall conductance at the critical disorder strength. This transition is similar to the one from the $f=1/3$ fermionic Laughlin state to an insulating phase reported in Refs.~\cite{Sheng03,Xin05,Zhao16}. However, such a similarity between bosons and fermions seems to be absent at the MR filling. As reported in Ref.~\cite{Wei19}, a disorder-induced intermediate FQH phase for $f=5/2$ Coulomb-interacting fermions might exist. By contrast, we have checked $f=1$ bosons interacting via either the contact repulsion or the Coulomb potential, but do not find any clear signature of an intermediate phase. Instead, we identify only a single transition from the bosonic MR state to a trivial insulating phase. Promising candidates for the disorder-induced intermediate phase of $f=5/2$ fermions include a PH symmetric puddle structure composed of fermionic MR and anti-MR domains~\cite{Mross18,Chong18,Biao18,Wei19} and a PH symmetric state~\cite{Jolicoeur07,Zucker16,Ajit18,Rezayi21}. As the usual PH transformation is only defined for fermions, the absence of an intermediate phase for $f=1$ bosons in our models does not contradict with the possible existence of an intermediate PH symmetric phase for $f=5/2$ fermions.

The remainder of this paper is organized as follows. In Sec.~{\ref{m_and_m}}, we introduce our model and method in detail, including the many-body Hamiltonian, the disorder model, and the definitions of ground-state entanglement entropy, fidelity susceptibility, and Hall conductance which are used to identify the phase transitions.
In Sec.~{\ref{result}}, we discuss the results at $f=1/2$, which will also benchmark the validity of our method. Then we will move to $f=1$ in Sec.~{\ref{result2}}. Finally, our conclusions, discussions and outlook are presented in Sec.~{\ref{c_and_o}}.

\section{Model and method}\label{m_and_m}
\subsection{Model}
We consider $N$ bosons in a two-dimensional (2D) random potential $U(\bf{r})$ on an $L_1\times L_2$ rectangular torus penetrated by a uniform perpendicular magnetic field. Bosons interact with each other via a translationally invariant two-body interaction $V({\bf r})$. After using the magnetic length $\ell_B$ as the length unit and choosing the Landau gauge, we can write the LLL single-particle orbitals as
\begin{equation}\label{torus_wf}
\begin{aligned}
\psi_m({\bf r}) = \left(\frac{1}{\sqrt{\pi} L_{2}}\right)^{\frac{1}{2}} \sum_{n=-\infty}^{+\infty} &e^{i \frac{2 \pi}{L_{2}}\left(m+n N_{\phi}\right) y} \\
&\times e^{-\frac{1}{2}\left[x-\frac{2 \pi}{L_{2}}\left(m+n N_{\phi}\right)\right]^{2}},
\end{aligned}
\end{equation}
where ${\bf r}=(x,y)$ is the real-space coordinate, $N_\phi$ is the number of magnetic flux quanta penetrating the torus, and $m=0,\dots,N_\phi-1$ is the orbital index. As required by the magnetic translational invariance on the torus, we have $L_1 L_2=2\pi N_\phi$~\cite{Haldane85}. To approach the 2D limit and preclude other competing phases such as the stripe phase~\cite{stripe,Cooper2005,Seki2008}, we choose the square torus with $L_1=L_2=\sqrt{2\pi N_\phi}\equiv L$ throughout this paper. 

We focus on the strong-field situation such that both the interaction and disorder are small compared with the Landau level spacing. In this case, the system's Hamiltonian is $\sum_{i<j}^N V({\bf r}_i-{\bf r}_j)+\sum_{i=1}^N U({\bf r}_i)$ projected to the LLL, where ${\bf r}_i$ is the coordinate of the $i$th boson. 
In the Fock space spanned by the single-particle orbitals [Eq.~(\ref{torus_wf})], the LLL-projected Hamiltonian after second quantization takes the form of
\begin{equation}\label{Hamiltonian}
\begin{aligned}
H=&\sum_{m_{1}, m_{2}, m_{3}, m_{4}=0}^{N_{\phi}-1} V_{m_{1}, m_{2}, m_{3}, m_{4}} c_{m_{1}}^{\dagger} c_{m_{2}}^{\dagger} c_{m_{3}} c_{m_{4}}\\
&+\sum_{m_{1}, m_{2}=0}^{N_{\phi}-1} U_{m_{1}, m_{2}} c_{m_{1}}^{\dagger} c_{m_{2}},
\end{aligned}
\end{equation}
where $c^{\dag}_m$ ($c_m$) creates (annihilates) a boson in the LLL orbital $m$, and the interaction and disorder matrix elements are 
\begin{equation}\label{interaction_term}
\begin{aligned} 
V_{\left\{m_{i}\right\}}=& \frac{1}{2} \delta_{m_{1}+m_{2}, m_{3}+m_{4}}^{\bmod N_{\phi}} \sum_{s, t=-\infty}^{+\infty} \delta_{t, m_{1}-m_{4}}^{\bmod N_{\phi}} V_\mathbf{q} \\
& \times e^{-\frac{1}{2}|\mathbf{q}|^{2}} e^{i \frac{2 \pi s}{N_{\phi}}\left(m_{1}-m_{3}\right)}
\end{aligned}
\end{equation}
and
\begin{equation}\label{disorder_term}
U_{\left\{m_{i}\right\}}=\sum_{s, t=-\infty}^{+\infty} \delta_{t, m_{1}-m_{2}}^{\bmod N_{\phi}} U_\mathbf{q} e^{-\frac{1}{4} |\mathbf{q}|^{2}} e^{\mathrm{i} \frac{\pi s}{N_{\phi}}\left(2 m_{1}-t\right)},
\end{equation}
respectively. In Eqs.~(\ref{interaction_term}) and (\ref{disorder_term}), $\delta_{i,j}^{\bmod N_{\phi}}$ is the periodic Kronecker delta function with period $N_{\phi}$, ${\bf q}=(q_x,q_y)=(2\pi s/L_1,2\pi t/L_2)$ with $|{\bf q}|^2=q^2_x+q^2_y$, and $V_{\bf q}=\frac{1}{2\pi N_\phi}\int V({\bf r}) e^{-{\rm i}{\bf q}\cdot{\bf r}}d{\bf r}$ and $U_{\bf{q}}=\frac{1}{2\pi N_\phi}\int U({\bf r}) e^{-{\rm i}{\bf q}\cdot{\bf r}}d{\bf r}$ are the Fourier transforms of $V({\bf r})$ and $U({\bf r})$, respectively. To study the effects of correlated disorder, we use the Gaussian correlated random potential for $U({\bf r})$, which satisfies 
$\left\langle U(\mathbf{r})U\left(\mathbf{r}^{\prime}\right)\right\rangle=\frac{W^{2}}{2 \pi \xi^{2}} e^{-\frac{\left|\mathbf{r}-\mathbf{r}^{\prime}\right|^{2} }{ 2 \xi^{2}}}$
and
$\left\langle U_\mathbf{q} U_{\mathbf{q^\prime}}\right\rangle = \frac{W^2}{2\pi N_\phi}
\delta_{\mathbf{q},-\mathbf{q}^{\prime}} e^{-q^{2}\xi^{2}/2}$,
where $\xi$ is the characteristic correlation length, $W$ is the disorder strength, and $\left\langle\cdots\right\rangle$ represents the sample average. If $\xi=0$, we return to the Gaussian white noise. When averaging over $N_s$ samples, we estimate the error bar of quantity $A$ by $\sqrt{\left(\left\langle A^{2}\right\rangle-\langle A\rangle^{2}\right) /\left(N_{s}-1\right)}$.

\subsection{Entanglement entropy}\label{ee}
Topologically ordered phases are characterized by the underlying pattern of quantum entanglement~\cite{Eisert10, Kitaev06, Wen06}. Previous works have found that the ground-state entanglement entropy and its derivative with respect to the disorder strength provide sharp signatures of the disorder-driven phase transitions in FQH systems~\cite{Zhao16,Zhao17,Wei19}. Here, we will apply this diagnosis to bosonic systems. For all cases that we consider, the system at zero and weak disorder is in a topological FQH phase (the bosonic Laughlin state at $f=1/2$ and the bosonic MR state at $f=1$) with a well-defined ground-state manifold consisting of $D$ approximately degenerate states, which are separated by a finite energy gap from other highly excited states. This ground-state degeneracy is a character of the underlying FQH topological order~\cite{WenNiu90}. Therefore, it is natural to extract the entanglement entropy from this ground-state manifold at weak disorder. For consistency, we will always choose the lowest $D$ eigenstates $|\Psi_{i=1,\ldots,D}\rangle$ of the many-body Hamiltonian equation (\ref{Hamiltonian}) as the ground-state manifold and calculate its entanglement entropy at all disorder strengths, even if the quasidegeneracy among states in this manifold disappears after the FQH phase is destroyed by strong disorder. 

Having defined the ground-state manifold, we divide the whole system into two subsystems $A$ and $B$ to calculate the entanglement entropy between them. We consider two kinds of half-half bipartition: (i) the orbital cut~\cite{Haque07}, for which $A$ consists of orbitals $m=0,\dots,\left\lceil N_{\phi} / 2\right\rfloor-1$ while $B$ consists of the remaining orbitals, respectively, where $\left\lceil a\right\rfloor$ is the integer part of $a$; and (ii) the real-space cut~\cite{Sterdyniak12,Dubail12}, for which $A$ is the region with  $x\in[0,L/2],y\in[0,L]$ and $B$ is the complement of $A$. The entanglement entropy between $A$ and $B$ can be measured by the von Neumann entropy $S(\rho)=-\operatorname{Tr} \rho_{A} \ln \rho_{A}$, where $\rho$ is the density matrix of the ground-state manifold and $\rho_A =\operatorname{Tr}_B \rho$ is the reduced density matrix of part $A$. Here, $\rho$ can be chosen as either the average over all states in the ground-state manifold, i.e.,  $\bar{\rho}=\frac{1}{D} \sum_{i=1}^{D}\left|\Psi_{i}\right\rangle\left\langle\Psi_{i}\right|$, or a single state $\rho_{i}=\left|\Psi_{i}\right\rangle\left\langle\Psi_{i}\right|$. These two choices correspond to the entanglement entropy $S(\bar{\rho})$ and
$\bar{S}=\frac{1}{D} \sum_{i=1}^{D}S(\rho_i)$, respectively. Both $S(\bar{\rho})$ and
$\bar{S}$ include the contributions of all $|\Psi_i\rangle$'s, thus minimizing the finite-size effect. We have checked that they give similar results, but $\bar{S}$ suffers from larger finite-size effects. Therefore our discussion will be based on the results of $S(\bar{\rho})$ in what follows.

\subsection{Fidelity susceptibility}\label{fs}
We will also use the ground-state fidelity and fidelity susceptibility to identify the disorder-driven phase transitions. For a parameter-dependent Hamiltonian, fidelity is originally the overlap between two ground states at different values of parameters~\cite{Zanardi06}. In our case, it can be generalized to the total overlap between two ground-state manifolds at disorder strength $W$ and $W+\Delta W$, i.e.,
\begin{equation}\label{fidelity}
F(W,\Delta W)=\left(\frac{1}{D} \sum_{i,j=1}^{D}|\left\langle\Psi_i(W)|\Psi_j(W+\Delta W)\right\rangle|^2\right)^{1/2}.
\end{equation}
Since the fidelity measures the similarity between states, the change in the ground-state manifold at the phase transition should be reflected by a dramatic drop in the fidelity across the critical point. A closely related quantity is the fidelity susceptibility, defined as
\begin{equation}\label{susceptibility}
\chi(W)= \lim_{\Delta W\rightarrow0}\frac{-2\ln F(W,\Delta W)}{\Delta W^2},
\end{equation}
which should show a sharp peak diverging with the increasing system size at the phase transition and can describe the universality class of the transition~\cite{Gu07,Gu08}.

\subsection{Hall conductance}\label{hc}
The Hall conductance $\sigma_H$ can unambiguously distinguish FQH phases with nonzero quantized $\sigma_H$ from insulating phases with $\sigma_H=0$ that could exist at strong disorder~\cite{Sheng03,Xin05}. $\sigma_H$ is simply proportional to the many-body Chern number of the ground state~\cite{Niu85}. After imposing the twisted boundary condition $\mathcal{T}_{\mu}\psi_m({\bf r})=e^{i\theta_\mu}\psi_m({\bf r})$ on the single-particle-orbital equation (\ref{torus_wf}), where $\mathcal{T}_\mu$ is the magnetic translation operator~\cite{Haldane85} in the $\mu=x,y$ direction and $\theta_\mu$ is the boundary phase, we can calculate the many-body Chern number 
\begin{equation}\label{chern}
C_j=\frac{i}{2\pi}\iint_0^{2\pi}d\theta_x d\theta_y\left(\left\langle\frac{\partial\Psi_j}{\partial \theta_x}\bigg|\frac{\partial\Psi_j}{\partial \theta_y}\right\rangle
-\left\langle\frac{\partial\Psi_j}{\partial \theta_y}\bigg|\frac{\partial\Psi_j}{\partial \theta_x}\right\rangle\right)
\end{equation}
for the state $|\Psi_j\rangle$ in the ground-state manifold. 
Numerically, we separate the $\theta_x$-$\theta_y$ space into dense meshes and calculate the sum of the accumulated Berry phase (divided by $2\pi$) in $|\Psi_j\rangle$ along the edges of each mesh to get $C_j$. Then we compute the total many-body Chern number $C=\sum_{j=1}^D C_j$ of the ground-state manifold, which gives the average ground-state Hall conductance $\sigma_H=(C/D) q^2/h$ with $q$ being the boson's charge. For an FQH phase at filling $f$, we expect $C=fD$ and $\sigma_H=f q^2/h$. For an insulating phase, we expect $C=0$ and $\sigma_H=0$.

\section{$f=1/2$ bosonic Laughlin filling}
\label{result}
Let us first study the effect of disorder on the $f=1/2$ bosonic Laughlin state to examine the validity of the diagnoses proposed in Secs.~\ref{ee}--\ref{hc}. In this section we assume that the bosons interact via the two-body contact repulsion with constant $V_{\bf q}=2/N_\phi$ in Eq.~(\ref{interaction_term}), i.e., the zeroth Haldane's pseudopotential~\cite{Haldane83}. In this case, the ground state in the absence of disorder is the model bosonic Laughlin state which is exactly two-fold degenerate on the torus. The corresponding wave function on an infinite plane~\cite{Laughlin83} takes the simple form 
\begin{equation}
\Psi^{f=1/2}_{\text{Laughlin}}=\prod_{j<k}\left(z_{j}-z_{k}\right)^{2} \exp \left[-\frac{1}{4} \sum_{i}\left|z_{i}\right|^{2}\right],
\end{equation}
where $z_j=x_j-iy_j$ is the complex coordinate of the $j$th boson on the plane. For the disorder term in Eq.~(\ref{disorder_term}), we first consider the Gaussian white noise. One can imagine that the ground state stays in the bosonic Laughlin phase when the disorder is weak but enters an insulating phase at strong disorder with all bosons pinned at the minimum of the disorder potential. Therefore at least one phase transition must occur at an intermediate disorder strength.

\begin{figure*}
	\centering
	\includegraphics[width=\linewidth]{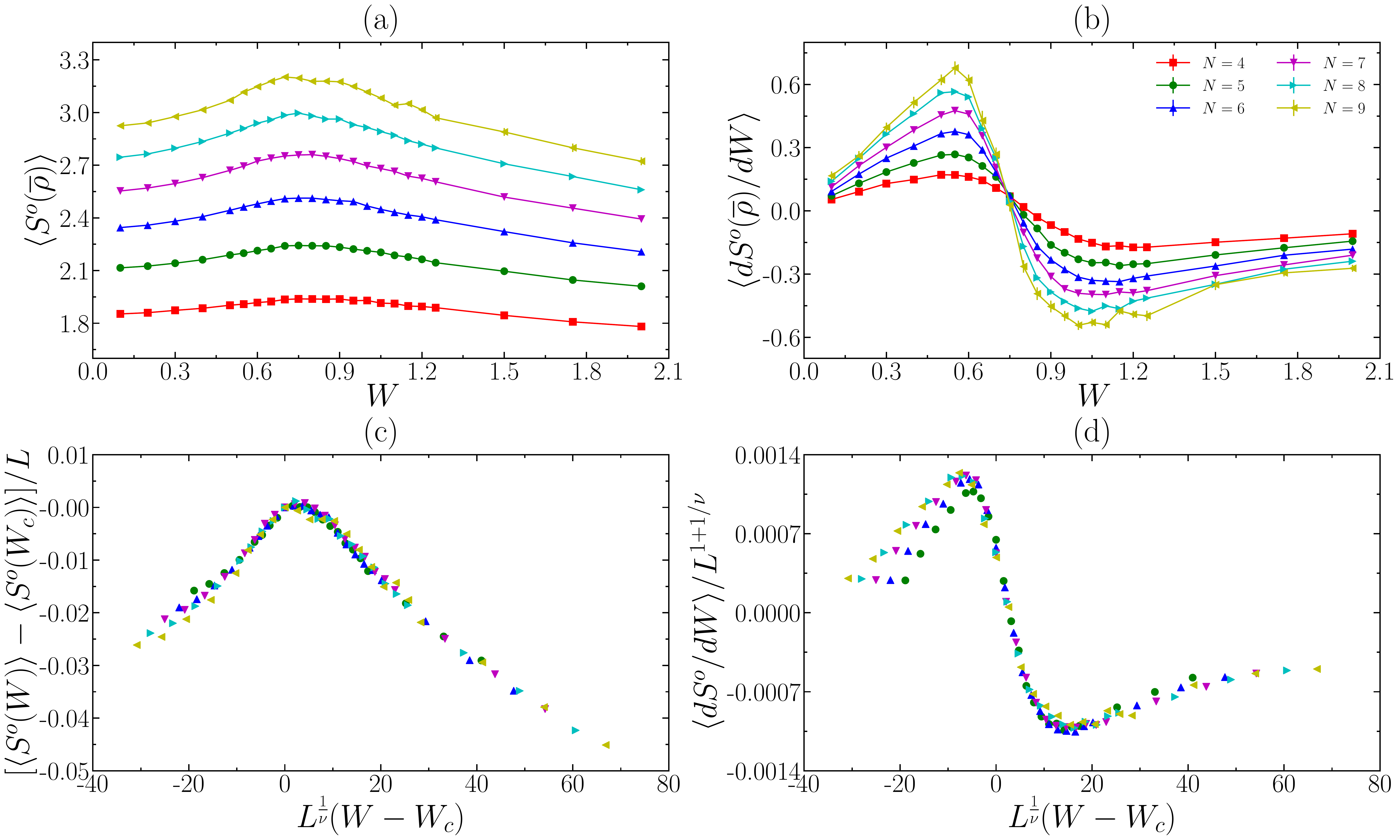}
	\caption{The orbital-cut entanglement entropy $S^o(\bar{\rho})$ for $N=4$-$9$ contact-interacting bosons at $f=1/2$. Disorder is modeled by the Gaussian white noise. (a) $\langle S^o(\bar{\rho})\rangle$ vs $W$. (b) $\langle dS^o(\bar{\rho})/dW\rangle$ vs $W$. For the finite-size scaling analysis of $\langle S^o(W)\rangle$ and $\langle dS^o/dW\rangle$, we plot $[\langle S^{o}(W)\rangle-\langle S^{o}(W_c)\rangle]/L$ and $\langle dS^o/dW\rangle/L^{1+1/\nu}$ vs $L^{1/\nu}(W-W_c)$ in (c) and (d), respectively, where we use $W_c=0.7$ and $\nu=0.6$ and neglect the smallest system size $N=4$. We average over 10 000 samples for $N=4$-$7$, 3500 samples for $N=8$, and 700 samples for $N=9$. Markers in (a)--(d) with the same color refer to the same system size.}
	\label{OEE_TwoBodyV0}
\end{figure*} 

To identify the phase transition, we measure the orbital-cut entanglement entropy (OEE), $S^o(\bar{\rho})$, as a function of disorder strength. At each $W$, we consider different disorder configurations and calculate the sample-averaged OEE $\left\langle S^o(\bar{\rho})\right\rangle$. Unlike for the $f=1/3$ fermionic Laughlin state, where $\left\langle S^o(\bar{\rho})\right\rangle$ monotonically decreases with increasing $W$ ~\cite{Zhao16,Zhao17}, here we find that $\left\langle S^o(\bar{\rho})\right\rangle$ first grows and then decays when disorder strength grows [Fig.~\ref{OEE_TwoBodyV0}(a)]. A peak in $\left\langle S^o(\bar{\rho})\right\rangle$ develops near $W=0.7$ for all system sizes, which seems to indicate a phase transition at $W\approx 0.7$. 

We further locate the critical point $W_c$ and extract the critical exponent $\nu$ of the correlation length via a scaling analysis of the OEE. As the OEE evolves smoothly with the disorder strength, we expect a continuous phase transition at $W_c$, with the correlation length 
\begin{equation}
\lambda \propto\left|W-W_{c}\right|^{-\nu}
\end{equation}
near the critical point, where $\nu$ is the critical exponent. Considering that the entanglement entropy obeys the area law, we argue the scaling form 
\begin{equation}
\label{oeescaling}
S^{o}(W)-S^{o}(W_c) \propto L g[L^{\frac{1}{\nu}}(W-W_c)]
\end{equation}
for the OEE, where $g$ is a universal function. By plotting the linear density of the OEE $[S^{o}(W)-S^{o}(W_c)]/L$ versus $L^{\frac{1}{\nu}}(W-W_c)$, we find that all data points in Fig.~\ref{OEE_TwoBodyV0}(a) (except those of the smallest system size $N=4$) collapse onto a single curve when we choose $W_c\approx0.7\pm 0.05$ and $\nu\approx0.6$ [Fig.~\ref{OEE_TwoBodyV0}(c)]. Such a scaling analysis provides a convincing indication of a phase transition at $W_c\approx0.7\pm 0.05$. Note that the critical exponent $\nu$ obtained here is close to the value for $f=1/3$ fermions~\cite{Zhao16}.

Compared with the entanglement entropy itself, its derivative with respect to the disorder strength may provide a sharper fingerprint of the underlying phase transition~\cite{Zhao16,Zhao17}. We approximate $dS^o(\bar{\rho})/dW$ in each sample as $[S^o(\bar{\rho})|_{W+\Delta W}- S^o(\bar{\rho})|_{W}]/\Delta W$ with $\Delta W=0.001W$, where $S^o(\bar{\rho})|_{W+\Delta W}$ is evaluated by only changing the magnitude of $W$ by a small percentage but keeping the disorder configuration fixed. The sample-averaged result is shown in Fig.~\ref{OEE_TwoBodyV0}(b). We observe one peak and one valley in $\langle dS^o(\bar{\rho})/dW\rangle$ when $\langle S^o(\bar{\rho})\rangle$ grows and decays with increasing $W$, respectively. The magnitudes of both the peak and valley grow with increasing system size. While it is difficult to confirm by our exact-diagonalization calculations in relatively small systems, we may expect that the peak and valley both move to the critical $W_c$ and their magnitudes diverge in the thermodynamic limit. In this case, $\langle dS^o(\bar{\rho})/dW\rangle$ goes to positive (negative) infinity when $W\rightarrow W_c$ from the left (right) side, which is a sharp signature of the phase transition. For finite systems, such divergence to positive and negative infinity is replaced by the peak and valley, respectively. Naively we can estimate $W_c$ as the position where $\langle dS^o(\bar{\rho})/dW\rangle\approx 0$, which again gives $W_c\approx 0.7$.  Moreover, according to Eq.~(\ref{oeescaling}), we suggest the scaling form
\begin{equation}
\label{dsdwscaling}
dS^{o}(W)/dW \propto L^{1+1/\nu} g^\prime [L^{\frac{1}{\nu}}(W-W_c)]
\end{equation}
to the entropy derivative, where $g^\prime$ means the derivative of the function $g$. In Fig.~\ref{OEE_TwoBodyV0}(d), we plot $\langle dS^{o}(W)/dW\rangle/L^{1+1/\nu}$ as a function of $L^{\frac{1}{\nu}}(W-W_c)$ for $W_c=0.7$ and $\nu=0.6$. Indeed, all data points in Fig.~\ref{OEE_TwoBodyV0}(b) (except those of the smallest system size $N=4$) collapse to a single curve near the critical point. 

At first glance, the non-monotonic behavior of the orbital-cut entanglement entropy in Fig.~\ref{OEE_TwoBodyV0}(a) seems counterintuitive. With increasing disorder strength, bosons are more likely trapped at the minima of the disorder potential, and thus we would expect weaker real-space entanglement between different regions of the system. Since the OEE is often argued to be a reasonable, approximated measure of the real-space entanglement between two subregions of the system due to the Gaussian localization of LLL orbitals~\cite{Haque07,LiH08}, one may expect a monotonically decaying OEE when disorder becomes stronger. However, this is contrary to our numerical observation that the OEE evolves non-monotonically with the disorder strength. We attribute this discrepancy to the imprecise capture of the real-space entanglement by the orbital cut: All LLL orbitals have nonzero support across the whole torus, so the orbital cut cannot precisely match the true real-space cut~\cite{Sterdyniak12,Dubail12}. Compared with fermions, this mismatch between orbital-cut and real-space entanglement may be more obvious for bosons due to their multiple occupation in each LLL orbital. Indeed, we did not observe the non-monotonic behavior of the OEE for fermions in previous works~\cite{Zhao16,Zhao17}. 

\begin{figure}
	\centering
	\includegraphics[width=\linewidth]{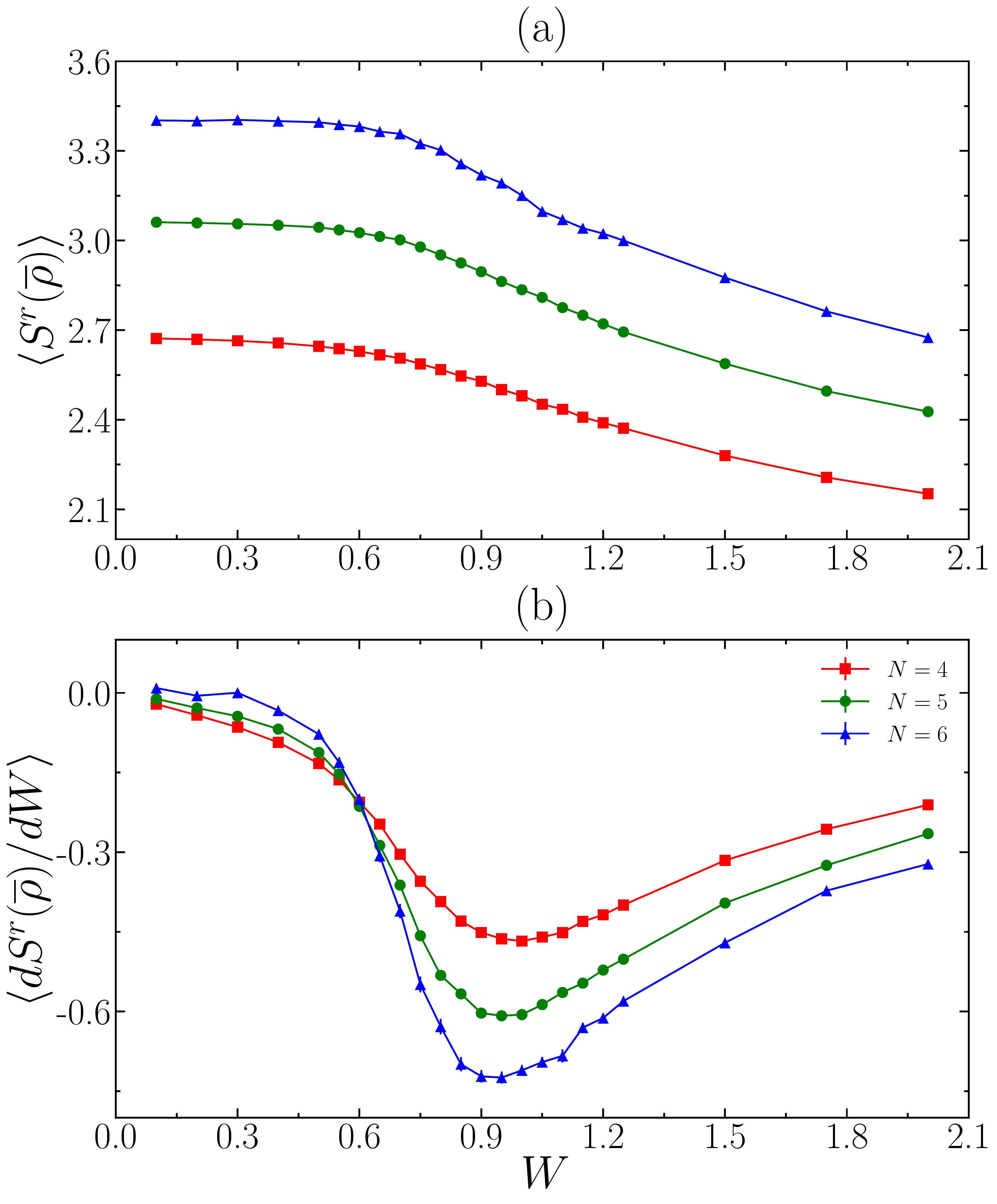}
	\caption{(a) The sample-averaged real-space-cut entanglement entropy $\langle S^r(\bar{\rho})\rangle$ and (b) the entropy derivative $\langle dS^r(\bar{\rho})/dW\rangle$ vs $W$ for $N=4$-$6$ contact-interacting bosons at $f=1/2$. Disorder is modeled by the Gaussian white noise. We average over 10 000 samples for $N=4$-$5$ and 1500 samples for $N=6$. Markers in (a) and (b) with the same color refer to the same system size.}
	\label{REE_TwoBodyV0}
\end{figure} 

In Fig.~\ref{REE_TwoBodyV0}(a), we show the real-space-cut entanglement entropy (REE) $S^r(\bar{\rho})$ obtained by a true real-space bipartition of the system instead of an orbital cut. Unlike $\langle S^o(\bar{\rho})\rangle$, $\langle S^r(\bar{\rho})\rangle$ indeed monotonically decays with increasing disorder strength, which is consistent with the picture that bosons are more likely trapped at minima of the disorder potential such that the REE becomes smaller. The derivative of $\langle S^r(\bar{\rho})\rangle$ with respect to the disorder strength exhibits a single minimum which gets deeper and flows to smaller $W$ for larger system sizes [Fig.~\ref{REE_TwoBodyV0} (b)]. Although we can only calculate the REE for at most six bosons because its computational cost is larger than that of the OEE~\cite{Sterdyniak12,Dubail12}, we still try a linear fitting of the minimum position of $\langle dS^r(\bar{\rho})/dW\rangle$ versus $1/N$ and obtain $W_c\approx0.78$ in the thermodynamic limit, which is consistent with the critical point indicated by the OEE. 

\begin{figure}
	\centering
	\includegraphics[width=\linewidth]{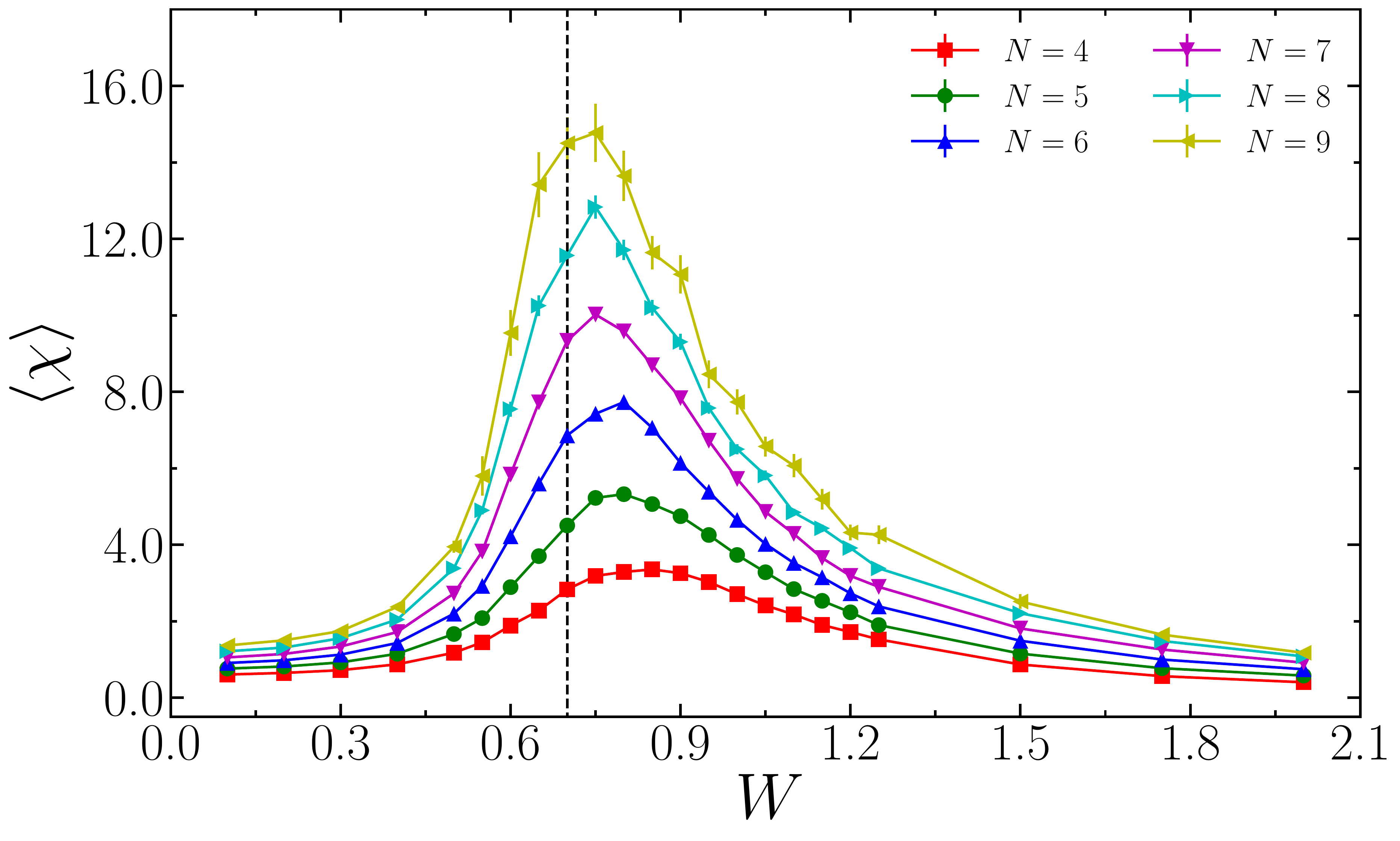}
	\caption{The sample-averaged fidelity susceptibility $\langle\chi(W)\rangle$ vs $W$ for $N=4$-$9$ contact-interacting bosons at $f=1/2$. Disorder is modeled by the Gaussian white noise. We average over 10 000 samples for $N=4$-$7$, 3500 samples for $N=8$, and 700 samples for $N=9$. The vertical dashed line indicates $W=0.7$, which is the critical disorder strength extracted from the entanglement quantities in Fig.~\ref{OEE_TwoBodyV0}.}
	\label{FS_TwoBodyV0}
\end{figure}

The diagnoses above based on entanglement measures suggest a single phase transition with increasing $W$. To confirm this, we look for a clue of the phase transition directly at the wave-function level. At each $W$, we fix the disorder configuration and change $W$ by $\Delta W=0.001W$ to evaluate the fidelity $F(W,\Delta W)$ in Eq.~(\ref{fidelity}). Then the fidelity susceptibility $\chi(W)$ is obtained from Eq.~(\ref{susceptibility}), which is further averaged over different disorder configurations. 
We find a single pronounced maximum in $\langle\chi(W)\rangle$ which grows for larger system sizes and probably diverges in the thermodynamic limit (Fig.~\ref{FS_TwoBodyV0}). Such a single peak indicates a single abrupt change in the ground-state manifold, i.e., a single phase transition. The location of the peak for the largest three system sizes gives a rough estimate of $W_c\approx0.75$, agreeing with the prediction of the entanglement analysis. 

\begin{figure}
	\centering
	\includegraphics[width=\linewidth]{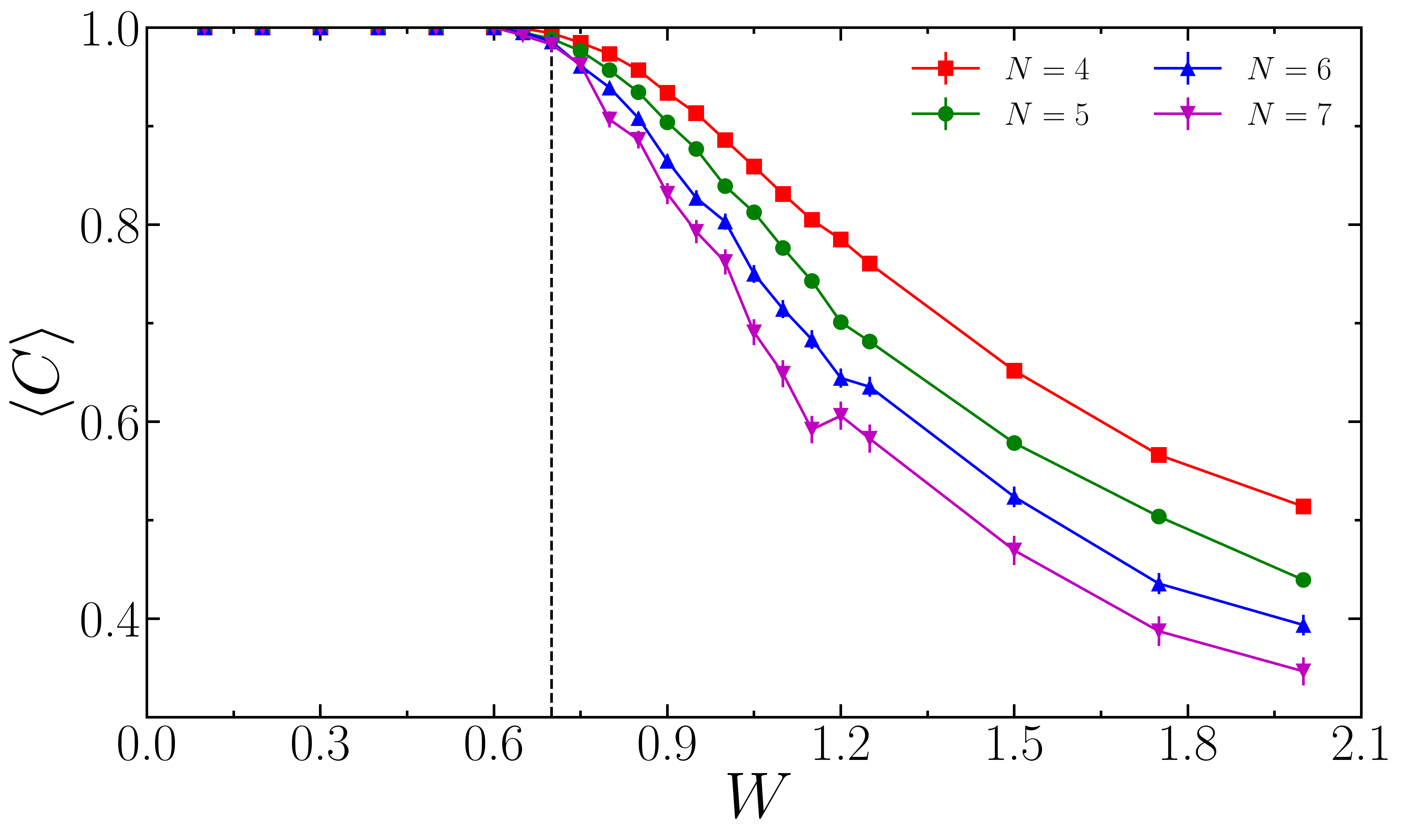}
	\caption{The sample-averaged many-body Chern number $\langle C\rangle$ vs $W$ for $N=4$-$7$ contact-interacting bosons at $f=1/2$. Disorder is modeled by the Gaussian white noise. We average over 10 000 samples for $N=4$-$5$, 3000 samples for $N=6$, and 1500 samples for $N=7$. The vertical dashed line indicates $W=0.7$, which is the critical disorder strength extracted from the entanglement quantities in Fig.~\ref{OEE_TwoBodyV0}.}
	\label{Chern_TwoBodyV0}
\end{figure}

\begin{figure*}
	\centering
	\includegraphics[width=\linewidth]{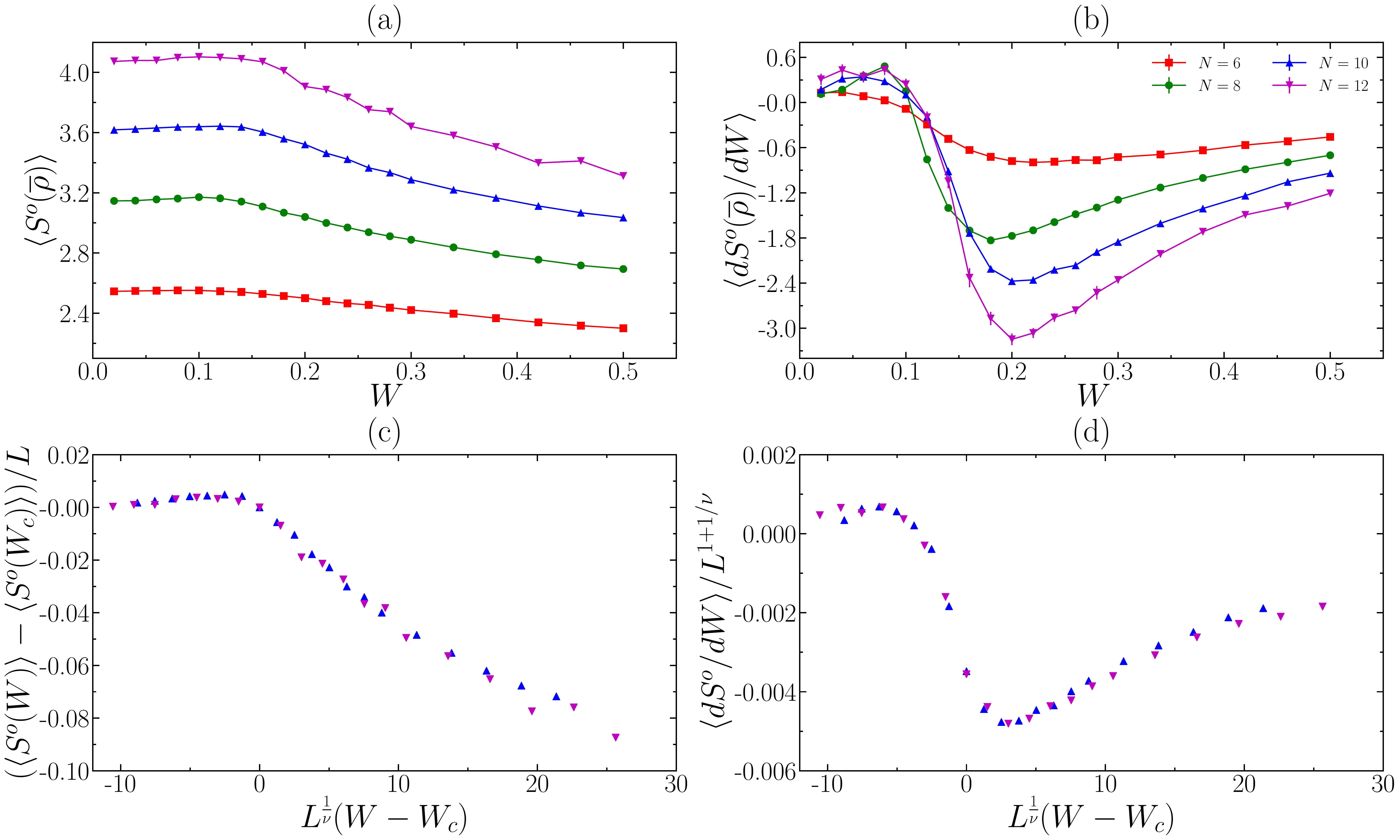}
	\caption{The orbital-cut entanglement entropy $S^o(\bar{\rho})$ for even $N=6$-$12$ Coulomb-interacting bosons at $f=1$. Disorder is modeled by the Gaussian white noise. (a) $\langle S^o(\bar{\rho})\rangle$ vs $W$. (b) $\langle dS^o(\bar{\rho})/dW\rangle$ vs $W$. For the finite-size scaling analysis of $\langle S^o(W)\rangle$ and $\langle dS^o/dW\rangle$, we plot $[\langle S^{o}(W)\rangle-\langle S^{o}(W_c)\rangle]/L$ and $\langle dS^o/dW\rangle/L^{1+1/\nu}$ vs $L^{1/\nu}(W-W_c)$ in (c) and (d), respectively, where we use $W_c=0.16$ and $\nu=0.5$. We average 10 000 samples for $N=6$ and $8$, 2000 samples for $N=10$, and 500 samples for $N=12$. Markers in (a)--(d) with the same color refer to the same system size.}
	\label{OEE_TwoBodyCoulomb_Thickness}
\end{figure*} 

In the end, we track the evolution of the many-body Chern number $C$ of the ground-state manifold as a function of disorder strength, which can unambiguously distinguish trivial insulators from FQH states. In the Laughlin phase, we should have $C=1$, while we expect $C=0$ in the insulating phase. For each disorder configuration, we twist the boundary conditions and evaluate $C$ by Eq.~(\ref{chern}). Then we average $C$ over various disorder configurations. Indeed, we observe $\langle C\rangle=1$ for weak disorder strengths (Fig.~\ref{Chern_TwoBodyV0}). A decaying of $\langle C\rangle$ starts at $W\approx 0.7$, signaling the collapse of the Laughlin phase. The drop in $\langle C\rangle$ becomes steeper for larger system sizes, tending towards a step function in the thermodynamic limit. 
Note that this critical disorder strength obtained from the many-body Chern number is consistent with the values extracted from the entanglement analysis and fidelity susceptibility, thus validating the reliability of all these diagnoses.

Beside the Gaussian white noise, we have also checked the Gaussian correlated random potential with nonzero correlation length $\xi$ for $f=1/2$ contact-interacting bosons and obtain similar results (not shown here to avoid repetition).
  
\section{$f=1$ bosonic Moore-Read filling}
\label{result2}
Having diagnosed the disorder-driven collapse of the $f=1/2$ bosonic Laughlin state, we now study the fate of the $f=1$ bosonic MR state in the presence of disorder. This bosonic MR state can be chosen as the model MR state, whose wavefunction on an infinite plane is 
\begin{equation}
\Psi^{f=1}_{\text{MR}}=\operatorname{Pf}\left(\frac{1}{z_{i}-z_{j}}\right) \prod_{i<j}\left(z_{i}-z_{j}\right) \exp \left[-\frac{1}{4} \sum_{k}\left|z_{k}\right|^{2}\right],
\end{equation}
with ``Pf'' standing for Pfaffian. 
Alternatively, we can also choose the ground state of a two-body interaction which is in the same topological phase as the model MR state. Previously, the Coulomb interaction was used to study the disorder-driven phase transitions and the disorder-induced intermediate FQH phase for fermions at the MR filling $f=5/2$~\cite{Wei19}. Therefore, for a better comparison with fermions, in this section we also use a two-body interaction and focus on the disorder effect on its ground state in the $f=1$ bosonic MR phase. This choice also reduces the computational cost, because obtaining the model bosonic MR state requires the numerical diagonalization of the three-body contact repulsion for an even number of bosons~\cite{Greiter91}.

On the torus geometry, the model bosonic MR state is three-fold degenerate in the $\mathbf{K}=(0, 0)$, $\mathbf{K}=(0, N/2)$, and $\mathbf{K}=(N/2, 0)$ momentum sectors~\cite{Cecile2015}, respectively. We hence need to seek a two-body interaction whose lowest three eigenstates are (i) in the same momentum sectors as the model MR states, (ii) approximately three-fold degenerate, (iii) separated from higher-energy levels by a robust energy gap, and (iv) close to the model MR states. One natural candidate for such a two-body interaction is the contact repulsion used in Sec.~\ref{result}, which is realistic for bosons in cold-atom setups. However, we observe a strong finite-size effect for this type of interaction. For relatively small systems of $N=6,8,10,12$ bosons, although the ground states of the two-body contact repulsion are in the same momentum sectors as the model bosonic MR states and the overlaps with the model states are decent, the energy gap protecting the ground-state manifold strongly fluctuates with the system size and can even become smaller than the splitting between the three ground states (see Table~\ref{MR_table}). Considering that we can only efficiently deal with at most $N=12$ bosons at $f=1$ once disorder is turned on, the strong finite-size effect of the two-body contact repulsion in small clean systems will lead to messy results at nonzero disorder. Indeed, in this case we find it very difficult to estimate $W_c$ and $\nu$, as shown in Appendix~\ref{result3}. 

\begin{table}
\caption{The ground-state energy gap $\Delta$ and the ground-state splitting $\delta$ at zero disorder for $f=1$ bosons. We consider both the two-body contact repulsion and the Coulomb interaction corrected by a finite layer thickness $w=4\ell_B$. We set $q^2/(4\pi\epsilon\ell_B)=1$ for the Coulomb interaction, where $\epsilon$ is the dielectric constant. After sorting all energy levels in ascending order, $\Delta$ is defined as the energy difference between the fourth level and the third level, and $\delta$ is the energy difference between the third level and the lowest level. For both interactions, the lowest three eigenstates are always in the same momentum sectors as the model bosonic MR state, forming the ground-state manifold. The total overlaps $\mathcal{O}=\sum_{i=1}^3|\langle\Psi_i|\Psi_i^{\mathrm{MR}}\rangle|^{2}/3$ between the ground-state manifold $\{|\Psi_i\rangle\}$ and the model bosonic MR states $\{|\Psi_i^{\mathrm{MR}}\rangle\}$ are also given. For $N=8$ and $N=10$ contact-interacting bosons, although the ground-state overlap with the model MR state is decent, $\Delta$ drops a lot and becomes smaller than $\delta$, implying that the ground-state manifold of these two systems is not in a robust MR phase.}
\begin{ruledtabular}
\begin{tabular}{lcccccc}
	& \multicolumn{3}{c} {Contact} & \multicolumn{3}{c} {\text{Coulomb with $w=4\ell_B$}} \\
	\hline
	& $\Delta$ & $\delta$ & $\mathcal{O}$ & $\Delta$ & $\delta$ & $\mathcal{O}$\\
	$N=6$  & 0.2531 & 0.0997 & 0.9582 & 0.0758 & 0.0356 & 0.9648\\
	$N=8$  & 0.0864 & 0.2028 & 0.8939 & 0.0443 & 0.0212 & 0.9676\\
	$N=10$ & 0.0928 & 0.1374 & 0.8968 & 0.0663 & 0.0212 & 0.9412\\
	$N=12$ & 0.3328 & 0.0879 & 0.8080 & 0.1190 & 0.0114 & 0.9109\\
	$N=14$ & 0.2771 & 0.1252 & 0.8152 & 0.1010 & 0.0206 & 0.9018\\
	$N=16$ & 0.3074 & 0.0420 & 0.7215 & 0.1251 & 0.0026 & 0.8749\\
\end{tabular}
\end{ruledtabular}
\label{MR_table}
\end{table}

Because the two-body contact repulsion is not ideal for our purpose, we examine long-range interactions. Motivated by the fact that the $f=5/2$ fermionic MR phase can be significantly enhanced by the Coulomb interaction with a finite layer thickness $w\approx 4\ell_B$ caused by the quasi-2D nature of the system~\cite{Peterson08L}, we check the energy spectra and ground-state wave functions of this interaction with $w=4\ell_B$ for $f=1$ bosons. Although the Coulomb interaction is not as pertinent as the contact repulsion for bosons in cold-atom setups, we think it is a reasonable choice for theoretical interest, as it can provide higher-order Haldane's pseudopotentials~\cite{Haldane83}. In this case, with the $V_{\bf q}$ given in Ref.~\cite{Peterson08L}, we find that there are indeed three approximately ground states in the correct momentum sectors. Remarkably, compared with the two-body contact repulsion, the Coulomb ground states have higher overlaps with the model bosonic MR state. Moreover, the energy gap suffers from a much weaker finite-size effect and is always larger than the ground-state splitting (Table~\ref{MR_table}). We hence proceed to study the disorder effect on the $f=1$ bosonic MR state stabilized by the the Coulomb interaction with $w=4\ell_B$. We emphasize that the trick of using the Coulomb interaction is only for suppressing the finite-size effect in numerically tractable systems. The results of both contact and Coulomb interactions are qualitatively similar, so the underlying physics should not depend on the choice of the interaction.

\begin{figure}
	\centering
	\includegraphics[width=\linewidth]{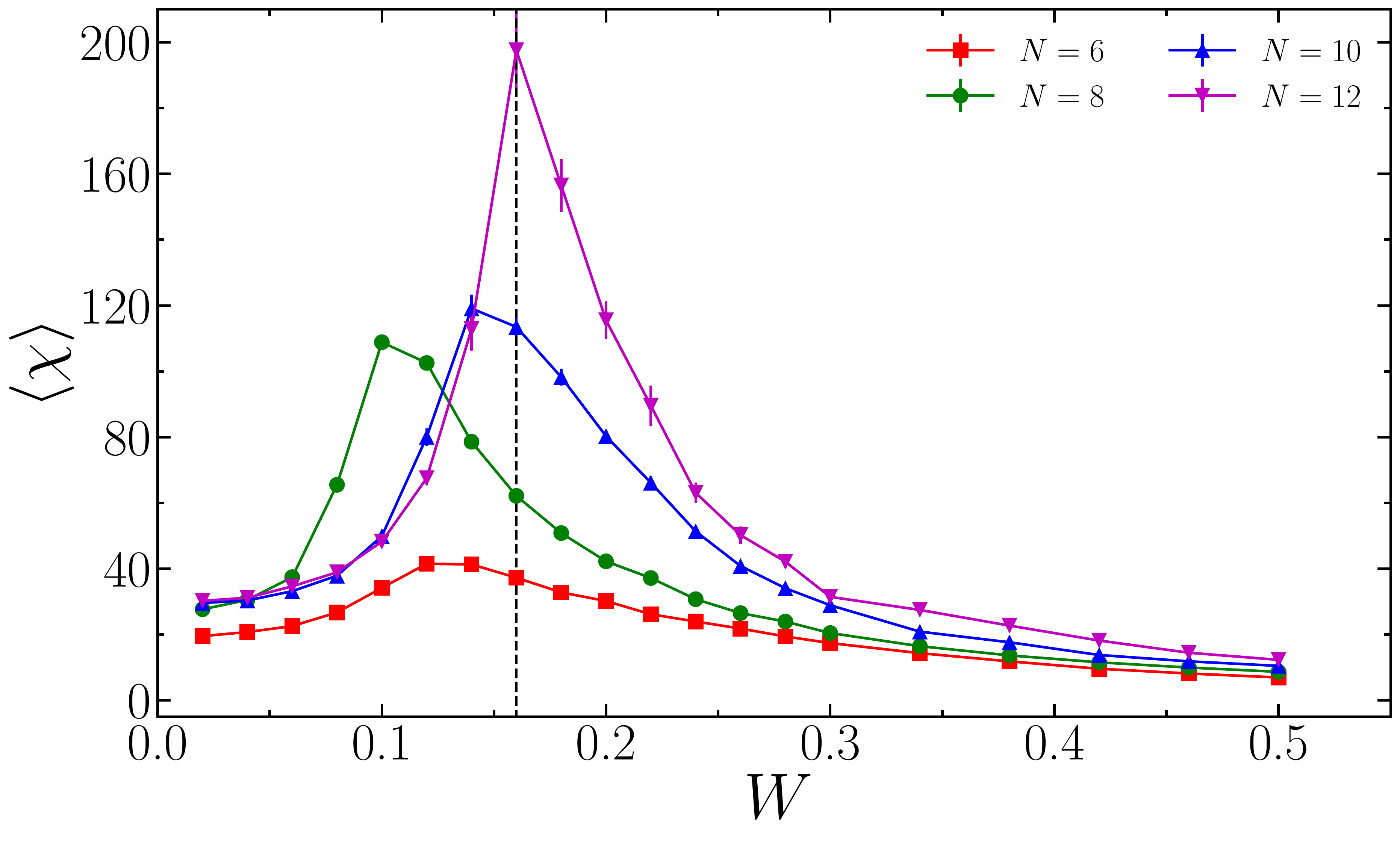}
	\caption{The sample-averaged fidelity susceptibility $\langle\chi(W)\rangle$ for even $N=6$-$12$ Coulomb-interacting bosons at $f=1$. Disorder is modeled by the Gaussian white noise. We average 10 000 samples for $N=6$ and $8$, 2000 samples for $N=10$, and 500 samples for $N=12$. The vertical dashed line indicates $W=0.16$, which is the critical disorder strength extracted from the entanglement quantities in Fig.~\ref{OEE_TwoBodyCoulomb_Thickness}.}
	\label{FS_TwoBodyCoulomb_Thickness}
\end{figure}

\begin{figure*}
 	\centering
 	\includegraphics[width=\linewidth]{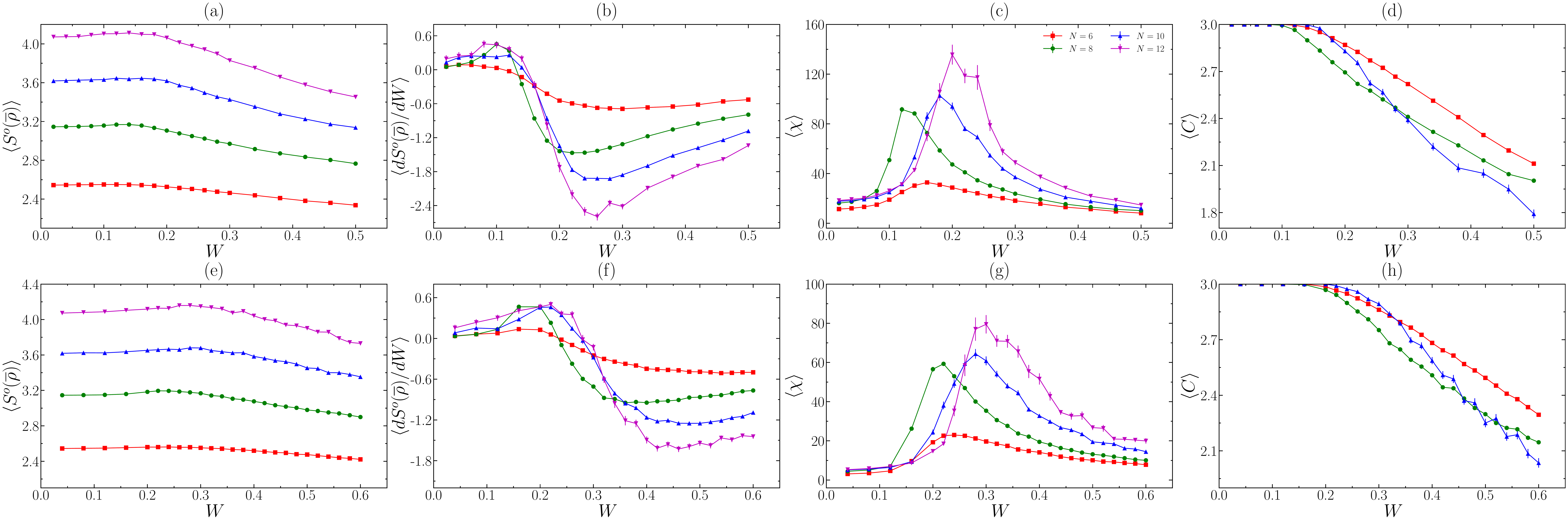}
 	\caption{The sample-averaged orbital-cut entanglement entropy $\langle S^o(\bar{\rho})\rangle$, entropy derivative $\langle dS^o(\bar{\rho})/dW\rangle$, fidelity susceptibility $\langle\chi(W)\rangle$, and many-body Chern number $\langle C\rangle$ for even $N=6$-$12$ Coulomb-interacting bosons at $f=1$. Disorder is modeled by the Gaussian correlated random potential with $\xi=0.5\ell_B$ [(a)--(d)] and $\xi=\ell_B$ [(e)--(h)]. In (a)--(c) and (e)--(g), we average 10 000 samples for $N=6$ and $8$, 2000 samples for $N=10$, and 500 samples for $N=12$. Due to the larger computational cost of the many-body Chern number, which requires integrating over boundary conditions, we average 10 000 samples for $N=6$, 5000 samples for $N=8$, and 1000 samples for $N=10$ in (d) and (h). Markers in (a)--(h) with the same color refer to the same system size.}
 	\label{CorrelatedDisorderCoulombThickness}
 \end{figure*} 

Now we switch on disorder. One can imagine that the system must enter an insulating phase at sufficiently strong disorder as all bosons are pinned at the minimum of the disorder potential. Therefore we expect at least one phase transition from the bosonic MR state to an insulating phase. Note that the $f=5/2$ fermionic MR or anti-MR state is ultimately replaced by a composite-fermion liquid rather than an insulator~\cite{Wei19}. This difference is due to the different statistics between bosons and fermions. In the presence of disorder, each Landau level is broadened into a Landau band, and all single-particle states except those at the band center are localized. While in the strong-disorder limit all bosons can pile in one localized single-particle state with the lowest energy, $f=5/2$ fermions must occupy half of all single-particle states in the second Landau band including the delocalized states at the band center, as two fermions cannot occupy the same single-particle state. 

To proceed, we first model disorder by the Gaussian white noise. Similar to the Laughlin case, the OEE $\left\langle S^o(\bar{\rho})\right\rangle$ first increases and then decreases with increasing $W$ for all system sizes [Fig.~\ref{OEE_TwoBodyCoulomb_Thickness}(a)], leading to a peak and a valley in the entropy derivative $\left\langle dS^o(\bar{\rho})/dW\right\rangle$ [Fig.~\ref{OEE_TwoBodyCoulomb_Thickness}(b)]. By contrast, we find that the REE decays monotonically (not shown here). We believe that the critical point $W_c$ can be identified by maximum of the OEE, and the OEE derivative diverges to positive (negative) infinity on the left (right) side of the critical point in the thermodynamic limit. Because the OEE data in Fig.~\ref{OEE_TwoBodyCoulomb_Thickness}, especially the entropy derivative, suffer from a stronger finite-size effect than the $f=1/2$ Laughlin case, we use the data of the largest two system sizes, $N=10$ and $N=12$, to do the scaling analysis to extract $W_c$. We apply the scaling forms in Eqs.~(\ref{oeescaling}) and (\ref{dsdwscaling}) to $\left\langle S^o(\bar{\rho})\right\rangle$ and $\left\langle dS^o(\bar{\rho})/dW\right\rangle$, respectively. With the rescaled variables, we find that all data points almost locate on a single curve with $W_c\approx0.16$ and $\nu\approx0.5$ [Figs.~\ref{OEE_TwoBodyCoulomb_Thickness}(c) and~\ref{OEE_TwoBodyCoulomb_Thickness}(d)].

To examine the critical $W$ extracted from the entanglement quantities, we calculate the fidelity susceptibility $\langle\chi(W)\rangle$ (Fig.~\ref{FS_TwoBodyCoulomb_Thickness}). Similar to the Laughlin case, all $\langle\chi(W)\rangle$ curves exhibit a pronounced maximum which gets higher for larger system size. While the position of the peak varies with the system size, it is located at $W\approx0.16$ for the largest system, $N=12$, which matches the $W_c$ extracted from the OEE. 

The similar behavior of the OEE to the $f=1/2$ Laughlin case and the single peak in the fidelity susceptibility suggest a direct phase transition from the bosonic MR state to an insulating phase when the disorder is modeled by the Gaussian white noise. Motivated by the intriguing observation that $f=5/2$ fermions subjected to disorder with a finite correlation length $\xi$ might evolve to an intermediate FQH phase~\cite{Wei19}, we repeat the studies in Figs.~\ref{OEE_TwoBodyCoulomb_Thickness} and \ref{FS_TwoBodyCoulomb_Thickness} by using the Gaussian correlated random potential with nonzero $\xi$. We consider $\xi/\ell_B=0.5$-$2$, for which a disorder-induced intermediate FQH phase seems to exist for $f=5/2$ fermions~\cite{Wei19}. However, for $f=1$ bosons, we cannot see any clear signature of an intermediate phase within this range of $\xi$. The results for $\xi=0.5\ell_B$ and $\xi=\ell_B$ are shown in Fig.~\ref{CorrelatedDisorderCoulombThickness}. In these cases, the OEE and the fidelity susceptibility look qualitatively similar to the $\xi=0$ case (i.e., the Gaussian white noise). While the curves at $\xi>0$ are stretched compared with the $\xi=0$ case such that the critical $W$ moves to larger values [Figs.~\ref{CorrelatedDisorderCoulombThickness}(a)--\ref{CorrelatedDisorderCoulombThickness}(c), \ref{CorrelatedDisorderCoulombThickness}(e)--\ref{CorrelatedDisorderCoulombThickness}(g)], we can only observe the signature of a single phase transition, as reflected by the single pronounced peak in the fidelity susceptibility [Figs.~\ref{CorrelatedDisorderCoulombThickness}(c) and~\ref{CorrelatedDisorderCoulombThickness}(g)]. This is also true for larger $\xi$ despite stronger finite-size effects. We also find that the many-body Chern number stays at $\langle C\rangle=3$ at weak disorder and decays at sufficiently large $W$ [Figs.~\ref{CorrelatedDisorderCoulombThickness}(d) and~\ref{CorrelatedDisorderCoulombThickness}(h)], which confirms the transition from the bosonic MR state to an insulating phase.

\begin{figure*}
\centering
\includegraphics[width=\linewidth]{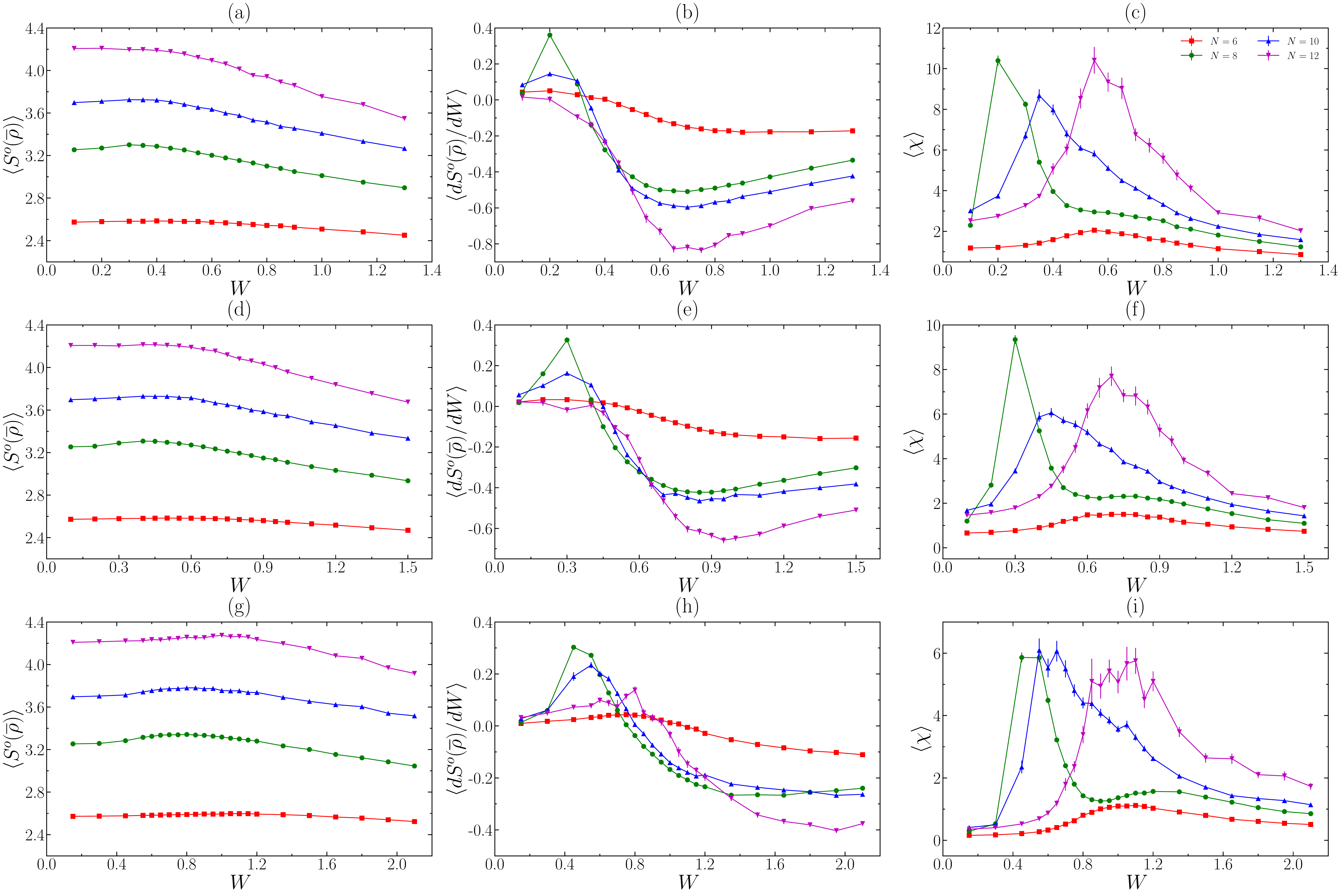}
\caption{The sample-averaged orbital-cut entanglement entropy $\langle S^o(\bar{\rho})\rangle$, entropy derivative $\langle dS^o(\bar{\rho})/dW\rangle$, and fidelity susceptibility $\langle\chi(W)\rangle$ for even $N=6$-$12$ contact-interacting bosons at $f=1$. Disorder is modeled by the Gaussian correlated random potential with $\xi=0$ [(a)--(c)], $\xi=0.5\ell_B$ [(d)--(f)], and $\xi=\ell_B$ [(g)--(i)]. We average 10 000 samples for $N=6$ and $8$, 2000 samples for $N=10$, and 500 samples for $N=12$. Markers in (a)--(i) with the same color refer to the same system size.}
\label{CorrelatedDisorderV0}
\end{figure*} 

\section{Conclusions and Discussions}\label{c_and_o}
In this paper, we present a systematic exact-diagonalization study of disorder-driven quantum phase transitions for the $f=1/2$ and $f=1$ bosonic FQH systems. We identify the phase transition by tracking the entanglement entropy, fidelity susceptibility, and many-body Chern number (Hall conductance) of the ground-state manifold as a function of disorder strength. The critical disorder strengths $W_c$ obtained from different quantities are consistent with each other, validating the reliability of our numerical results. At both $f=1/2$ and $f=1$, we identify a single phase transition from the FQH state (bosonic Laughlin state for $f=1/2$ and bosonic MR state for $f=1$) to an insulating phase, no matter whether the correlation length in the disorder potential is zero or not. While the location of the transition appears to be robust, the critical exponent $\nu$ of the transition extracted from the finite-size scaling analysis of the OEE data ($\nu\approx0.6$ at $f=1/2$ and $\nu\approx 0.5$ at $f=1$) violates the expected Harris-Chayes inequality $\nu\geq 2/d$ for $d$-dimensional disordered systems~\cite{Harris1974,Chayes86}. Such violation was also observed in fermionic FQH systems~\cite{Zhao16,Zhao17} and many-body localization transitions~\cite{Kjall14,Luitz15}. Our results may motivate further studies of the critical exponent of disorder-driven phase transitions in FQH systems by more powerful numerical techniques and experiments. 

The fate of the $f=1$ bosonic MR state in the presence of disorder looks different from that of the $f=5/2$ fermionic MR or anti-MR state. At least for the interactions and the range of the disorder correlation length considered by us ($\xi/\ell_B=0.5$-$2$), we observe a direct transition to an insulating phase for the former, while for the latter the system might first enter an intermediate FQH phase before finally becoming a composite-fermion liquid~\cite{Wei19}. The suspected intermediate phase in the fermionic case was argued to be a PH symmetric phase described by either the fermionic-MR--anti-MR puddle structure~\cite{Mross18,Chong18,Biao18,Wei19} or a PH symmetric wavefunction~\cite{Jolicoeur07,Zucker16,Ajit18,Rezayi21}. 
Since the usual PH transformation is only well defined for fermions, it is indeed impossible for bosons to form a PH symmetric intermediate phase induced by disorder. Therefore our results do not contradict with the existence of a PH symmetric intermediate phase of $f=5/2$ fermions. In Ref.~\cite{Chong16}, a generalized PH transformation is defined for bosons, by which one can construct a bosonic anti-MR state at $f=1$. However, unlike in the fermionic case where the MR and anti-MR states have the same energies for PH symmetric Hamiltonians, this bosonic anti-MR state could have a much higher energy than the bosonic MR state. Indeed, the absence of an intermediate phase in our numerical results seems to suggest that the bosonic anti-MR state is not energetically favored by the Hamiltonian we use.

A future direction following our work could be exploring the disorder effect on dipolar interacting bosons, for which the FQH states compete with stripe and bubble phases~\cite{Cooper2005,Seki2008}. Alternatively, one can also consider disorder-driven phase transitions when the MR state competes with the composite-fermion liquid for bosons~\cite{Wu15}. It would be interesting to check whether an intermediate phase can appear in these cases. Another future direction could be to study the disorder-driven phase transitions for bosonic FQH states in optical lattice setups, i.e., the bosonic fractional Chern insulators~\cite{parameswaran2013fractional,liu2013review,Titusreview}. In that case, the notable differences between a Chern band and the LLL, such as the multiband structure and high Chern number, may host new interesting phenomena. 

\acknowledgements
We thank Bo Yang, Wei Zhu, and Xin Wan for fruitful discussions. This work is supported by the National Natural Science Foundation of China through Grant No.~11974014.

\appendix
\section{$f=1$ with the contact interaction}
\label{result3}

Here, we study the disorder-driven phase transitions for contact-interacting bosons at $f=1$. In Fig.~\ref{CorrelatedDisorderV0}, we present the orbital-cut entanglement entropy, the entropy derivative, and the fidelity susceptibility for different correlation lengths of disorder. Compared with Figs.~\ref{OEE_TwoBodyCoulomb_Thickness}--\ref{CorrelatedDisorderCoulombThickness}, the results with the contact interaction are qualitatively similar to the Coulomb-interacting case. Overall, there is no sign of an intermediate FQH phase. However, the data in Fig.~\ref{CorrelatedDisorderV0} are much messier than those in Figs.~\ref{OEE_TwoBodyCoulomb_Thickness}--\ref{CorrelatedDisorderCoulombThickness}. As we argue in the main text, this is due to the strong finite-size effect of the energy gap in small clean systems of $N=6,8,10,12$. For these system sizes, once we turn on disorder, the critical strength of disorder required to destroy the MR state strongly fluctuates with the system size. Even with the largest two systems, $N=10$ and $N=12$, we find it difficult to make a reliable scaling analysis for the entanglement entropy and its derivative to extract $W_c$ and $\nu$. Moreover, with increasing system size, the peak position of the fidelity susceptibility changes a lot, and the peak height does not monotonically grow, making it difficult to estimate $W_c$. All these results suggest that much larger system sizes are needed to study the disorder-driven phase transitions for contact-interacting bosons at $f=1$. Alternatively, one can choose the Coulomb interaction to suppress the finite-size effect, as done in the main text.



\bibliography{Disorder}

\end{document}